\documentclass[letterpaper, 10 pt, conference]{ieeeconf}  

\IEEEoverridecommandlockouts                              
\usepackage{graphicx}
\usepackage{subfig}
\usepackage[belowskip=-10pt,aboveskip=0pt]{caption}
\usepackage{amsmath} 
\usepackage{amssymb}  
\usepackage{acronym}
\usepackage{color}
\usepackage{gensymb}
\usepackage{authblk}
\usepackage{cite}
\usepackage{url}
\expandafter\def\expandafter\UrlBreaks\expandafter{\UrlBreaks
  \do\a\do\b\do\c\do\d\do\e\do\f\do\g\do\h\do\i\do\j%
  \do\k\do\l\do\m\do\n\do\o\do\p\do\q\do\r\do\s\do\t%
  \do\u\do\v\do\w\do\x\do\y\do\z\do\A\do\B\do\C\do\D%
  \do\E\do\F\do\G\do\H\do\I\do\J\do\K\do\L\do\M\do\N%
  \do\O\do\P\do\Q\do\R\do\S\do\T\do\U\do\V\do\W\do\X%
  \do\Y\do\Z}
  
\usepackage{tikz}
\usetikzlibrary{shapes.geometric, arrows}

\acrodef{LTE}{Long Term Evolution}
\acrodef{RACH}{Random Access Channel}
\acrodef{UE}{User Equipment}
\acrodef{BS}{Base Station}
\acrodef{ZC}{Zadoff-Chu}
\acrodef{DFT}{Discrete Fourier Transform}
\acrodef{LoS}{Line-of-Sight}
\acrodef{NLoS}{Non-Line-of-Sight}
\acrodef{MIMO}{Multiple Input Multiple Output}
\acrodef{OFDM}{Orthogonal Frequency Division Multiplexing}
\acrodef{SNR}{Signal to Noise Ratio}
\acrodef{CDF}{Cumulative Distribution Function}
\acrodef{CFO}{Carrier Frequency Offset}
\acrodef{UL}{Uplink}
\acrodef{DL}{Downlink}
\acrodef{FFT}{Fast Fourier Transform}
\acrodef{IFFT}{Inverse Fast Fourier Transform}
\acrodef{TDD}{Time Division Duplex}
\acrodef{FDD}{Frequency Division Duplex}
\acrodef{CE}{Channel Estimation}
\acrodef{SWP}{Spherical Wave Propagation}
\acrodef{SVD}{Singular Value Decomposition}
\acrodef{IMU}{Inertial Measurement Unit}
\acrodef{RB}{Resource Block}
\acrodef{HW}{Half Wave}

\title{\LARGE \bf
Full Downlink Channel Reconstruction using Incomplete Uplink Channel Measurements in Massive MIMO networks
}

\author[1]{Aleksei Fedorov}
\author[1]{Haibo Zhang}
\author[1]{Galina Sidorenko}
\author[2]{Bo Yang}
\affil[1]{Department of Computer Science,
			University of Otago, New Zealand}
\affil[2]{School of Electronic, Information and Electrical Engineering, Shanghai Jiao Tong University, China}

\begin{document}
	\def\spvec#1{\left(\vcenter{\halign{\hfil$##$\hfil\cr \spvecA#1;;}}\right)}
	\def\spvecA#1;{\if;#1;\else #1\cr \expandafter \spvecA \fi}

\tikzstyle{rect} = [draw, rectangle, fill=white!20,text width=20em, text centered, minimum height=1.5em]
\tikzstyle{rect_input} = [draw, rectangle, fill=gray!20,text width=20em, text centered, minimum height=1.5em]
\tikzstyle{line} = [draw, -latex']
\tikzstyle{cloud} = [draw, ellipse, fill=gray!20, node distance=3cm, text centered, minimum height=2em]

\maketitle
\thispagestyle{empty}
\pagestyle{empty}

\begin{abstract}
While more and more antennas are integrated into a single mobile user equipment to increase communication quality and throughput,  the number of antennas used for transmission is commonly restricted due to the concerns on hardware complexity and energy consumption, making it impossible to achieve the maximum channel capacity. This paper investigates the problem of reconstructing the full downlink channel from incomplete uplink channel measurements in Massive MIMO systems. 
We present ARDI, a scheme that builds a bridge between radio channel and physical signal propagation environment to link spatial information about the non-transmitting antennas with their radio channels. By inferring locations and orientations of the non-transmitting antennas from an incomplete set of uplink channels, ARDI can reconstruct the downlink channels for non-transmitting antennas. We derive closed-form solution to reconstruct antenna orientation in both single-path and multi-path propagation environments. The performance of ARDI is evaluated using simulations with realistic human movement.  The results demonstrate that ARDI is capable of accurately reconstructing full downlink channels when the signal-to-noise ratio is higher than 15dB, thereby expanding the channel capacity of Massive MIMO networks.


\end{abstract}

\newcommand{\vphi}{\varphi}
\newcommand{\bp}{\boldsymbol{p}}
\newcommand{\bq}{\boldsymbol{q}}
\newcommand{\be}{\boldsymbol{e}}
\newcommand{\bh}{\boldsymbol{h}}
\newcommand{\mPr}{\mathrm{Pr}}
\newcommand{\mPath}{\mathrm{Path}}
\newcommand{\bmPath}{\boldsymbol{\mathrm{Path}}}
\newcommand{\bV}{\boldsymbol{V}}
\newcommand{\bn}{\boldsymbol{n}}
\newcommand{\bE}{\boldsymbol{E}}
\newcommand{\bH}{\boldsymbol{H}}
\newcommand{\mNLoS}{\mathrm{NLoS}}
\newcommand{\mLoS}{\mathrm{LoS}}
\newcommand{\bxi}{\boldsymbol{\xi}}
\newcommand{\by}{\boldsymbol{y}}
\newcommand{\tUE}{\text{UE}}
\newcommand{\bw}{\boldsymbol{\omega}}
\newcommand{\bv}{\boldsymbol{v}}
\newcommand{\dd}{\!\!}
\newcommand{\td}{\!\!\!}
\newcommand{\MN}{M\! \times \!N}
\newcommand{\mN}{m\! \times \!N}

\newcommand{\sym}{\mathrm{sym}}
\newcommand{\Dt}{\Delta t}
\newcommand{\mI}{\mathrm{I}}

\section{Introduction}
The continuously increasing demand on high throughput wireless communication has forced the communication technology to integrate more and more antennas at both \ac{BS} and \ac{UE} sides to exploit the advantages of \ac{MIMO} to increase the capacity of the wireless channel. From the BS side, Massive MIMO, as one of the key technologies for 5G networks, 
tends to integrate even hundreds of antennas at one BS. Ericsson, Huawei, and Facebook have demonstrated Massive MIMO systems with as many as 96 to 128 antennas \cite{Ericsson, Huawei, Facebook}. From the UE side, the existing flagman smartphones such as Samsung S8, Note9, Sony XZ, Pixel 2 already have four antennas \cite{Samsung}. In 2018 Qualcomm unveils the first mmWave 5G antennas for smartphones, and its Snapdragon X50 modem can support up to 16 in one smartphone \cite{Qualcomm}. It is undoubted that more and more antennas will be added to both UEs and BSs in the near future.  

While the trend in increasing the number of antennas at the \ac{UE} side is evident, having more antennas for transmission will not only increase the hardware complexity and consume more energy, but also make the pilot contamination problem  even worse \cite{Marzetta}. Hence, the leading \ac{UE} producers are putting efforts to optimize the antenna design by limiting the number of antennas for transmission, that is, only use a subset of the antennas for transmission, and the others are receive-only. The \ac{UE} simply performs a weighted summation of signals from antennas with similar channels during signal reception. Since only a subset of antennas is used for transmission, the \ac{BS} inevitably measures an incomplete channel. Hence, the available channel capacity becomes smaller than the capacity of the full channel where all \ac{UE}'s antennas are involved in transmission, and the increase in throughput can be minuscule.

The question investigated in this work is: \textit{is it possible to reconstruct the full downlink channel between all antennas at a UE and a BS based on only the incomplete uplink channel measurements obtained from the subset of transmitting antennas at the UE?}  A possible approach is to use the frequency-independent reciprocity on propagation paths, which implies that \ac{UL} and \ac{DL} signals traverse through the same paths. This approach has been proposed to eliminate or significantly reduce the overhead caused by \ac{DL} channel estimation feedback \cite{MIT-Vasisht1, arXiv_DLCH_reconstruction}. Using the parameters of the propagation paths \cite{Lund_RIMAX, RIMAX_Transaction, My4, SAGE1}, the \ac{DL} channel can be inferred using \ac{UL} channel measurements. However, such an approach becomes inapplicable for receive-only antennas since the \ac{BS} cannot directly obtain the information about the propagation parameters for the receive-only antennas.

\begin{figure}[!ht]
	\centering
	\includegraphics[width=0.47\textwidth]{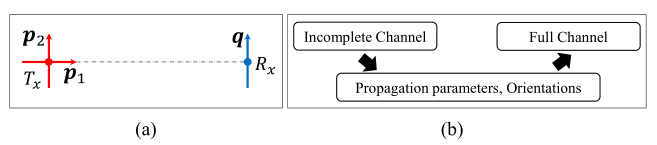}
	\caption{(a) Impact of the mutual orientations of the transmitting antenna ($T_x$) and the receiving antenna ($R_x$) on channel. (b) Reconstructing the full channel by inferring propagation parameters and orientations in ARDI.}
	\label{fig:Orientation}
\end{figure}

Since the form factor of a \ac{UE} is usually small especially for smartphones, it is commonly assumed that the channels for different antennas are similar \cite{My4}. However, as illustrated in Fig. \ref{fig:Orientation}(a), the channels for transmitting antennas with orientations $\bp_1$ and $\bp_2$ are significantly different. The reception from the antenna with orientation $\bp_1$ is almost equal to zero since it is perpendicular to the orientation of the receiving antenna $\bq$. This example shows that, even if the parameters of propagation paths are known, the channel for a receive-only antenna cannot be inferred due to unknown orientation, which can significantly affect channel reconstruction. 

\textbf{Novelty\&Contribution:} In this paper, we present ARDI (\textbf{A}ntenna orientation \textbf{R}econstruction and \textbf{D}ownlink channel \textbf{I}nference), a scheme that can reconstruct the full DL channel based on incomplete \ac{UL} channel measurements. 
The heart of ARDI is the reconstruction of UE antenna orientation at BS based on only \ac{UL} signals. We observe that the channel response is closely related to the mutual orientations of transmitting and receiving antennas. Based on this observation, we design an algorithm that allows the BS to use channel response as a measurement and calculate the orientation of transmitting antennas of UE based on the electromagnetic signal propagation model. Since the layout the antennas at a UE is known based on the UE models, the orientations and other propagation parameters for the non-transmitting antennas can then be inferred, based on which the full downlink channel can be reconstructed.  

To the best of our knowledge, ARDI is the first scheme that can reconstruct the orientation of an antenna based on a single electromagnetic impulse in both single-path and multipath propagation environments. We derived a closed-form solution for antenna orientation reconstruction in both propagation environments.  Although there are some works on antenna array orientation reconstruction \cite{orient1}, none of them can reconstruct the orientation of a single antenna or reconstruct orientation of a \ac{UE} based on two transmitting antennas.
Also, the proposed algorithm is the first that can reconstruct \ac{DL} channels for non-transmitting antennas. There are some existing works on \ac{DL} channel reconstruction \cite{MIT-Vasisht1, arXiv_DLCH_reconstruction}, but all of them consider only a single antenna at \ac{UE}s. We show that our scheme is suitable to both \ac{FDD} and \ac{TDD} modes, and demonstrate its performance through simulations.

\section{Overview of ARDI}\label{sec:overview}

\begin{figure}[!ht]
	\centering
	\includegraphics[width=2.5in]{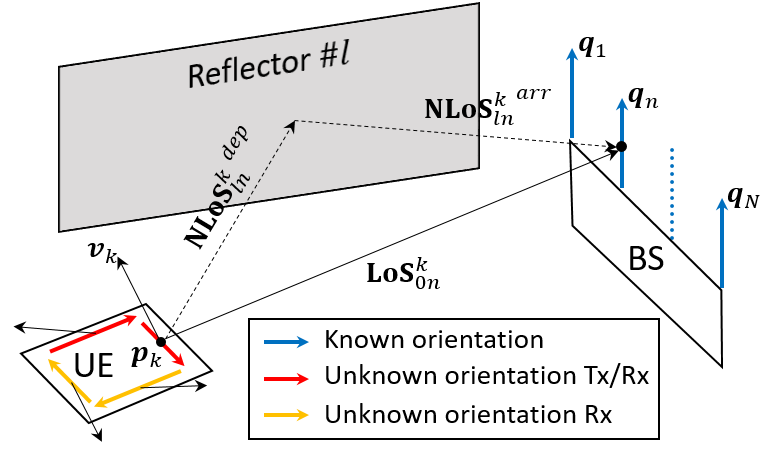}
	\caption{Communication between a UE and a BS.}
	\label{fig:multipath}
\end{figure}

As illustrated in Fig. \ref{fig:multipath}, we assume the BS has $N$ antennas, and the UE has $M$ antennas but only $m$ of them can transmit ($M=4$ and $m=2$ in the example given in Fig. \ref{fig:multipath}). We aim to increase the downlink channel capacity by expanding the downlink channel from $\mN$ to $\MN$. 

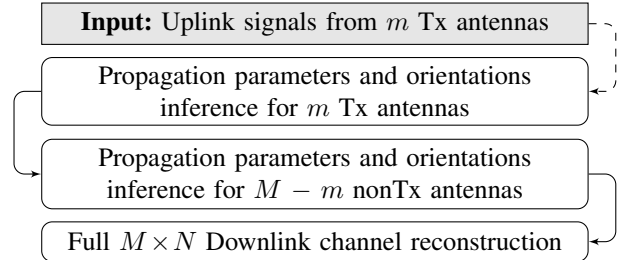
\begin{figure}[!ht]
\begin{center}
    \begin{tikzpicture}[node distance = 0.8cm, auto]
        \node [rect_input] (step1) {\textbf{Input:} Uplink signals from $m$ Tx antennas};
        \node [rect, rounded corners, below of=step1, node distance = 0.9cm] (step2) {Propagation parameters and orientations \\ inference for $m$ Tx antennas};
        \node [rect, rounded corners, below of=step2, node distance = 1.1cm] (step3) {Propagation parameters and orientations \\ inference for $M-m$ nonTx antennas}; 
        \node [rect, rounded corners, below of=step3, node distance = 0.9cm] (step4) {Full $M\!\times\! N$ Downlink channel reconstruction};
        \path [line, rounded corners, dashed] (step1) --++ (4,0) |- (step2);
        \path [line, rounded corners] (step2) --++ (-4,0) |- (step3);
        \path [line, rounded corners] (step3) --++ (4,0) |- (step4);
    \end{tikzpicture}
    \caption{Steps for reconstructing the full downlink channel.}
    \label{chart:ch_reconstruction}
\end{center}
\end{figure}

The \textbf{key idea} of ARDI is to reconstruct the full $\MN$ DL channel based on the spatial information such as propagation paths, locations, and orientations of the antennas inferred based on the UL signals from the transmitting antennas at UE to the antennas at the BS. Fig. \ref{chart:ch_reconstruction} shows the flowchart of ARDI. Firstly, the \ac{BS} estimates the propagation parameters of the transmitting antennas such as propagation paths, antenna location, and Doppler effect, and then it reconstructs orientations of the transmitting antennas based on the estimated propagation parameters. Based on the orientations and the propagation parameters, the \ac{BS} further infers the multipath propagation parameters and orientations for the non-transmitting antennas, and finally reconstruct the full $\MN$ DL channel based on the estimated propagation parameters. 



\textbf{Notations:} We use ($\cdot$) to denote the scalar product operation and ($\times$) to represent the vector product operation. The operation of transposition is represented with superscript $T$. $\|\cdot\|$ represents the $\ell^2$-norm, and bold letters represent vectors. 


\section{Antenna Orientation Reconstruction}\label{sec:antenna_orientation}

\begin{figure}[!ht]
	\centering
	\includegraphics[width=2.5in]{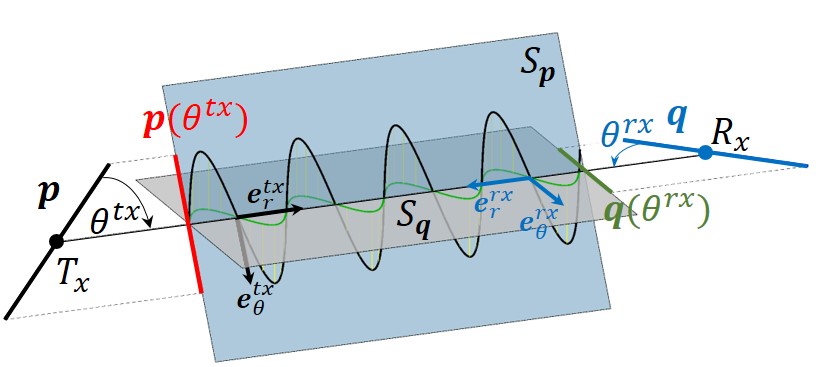}
	\caption{Definition of effective lengths and the electric field projection.}
	\label{fig:effective_length}
\end{figure}

In this section, we present the solution for the BS to reconstruct the orientation of a transmitting antenna of the UE based on only the measurements of the uplink signals. Our solution is motivated by the \textit{strong relation between the voltage induced at a receiving antenna and the mutual orientation of the transmitting and receiving  antennas}. We use Fig. \ref{fig:effective_length} to explain this relation. 

As shown in Fig. \ref{fig:effective_length}, a transmitting antenna centered at $T_x$ with orientation $\bp$ emits an electromagnetic signal that is received by a receiving antenna centered at $R_x$ with orientation $\bq$, where $\bp$ and $\bq$ are unit length vectors. Let $\overline{T_xR_x}$ be the line that connects the centers of the two antennas.  $S_{\bp}$ is the plane determined by $\bp$ and $\overline{T_xR_x}$, and $S_{\bq}$ is the plane determined by $\bq$ and $\overline{T_xR_x}$. $\theta^{tx}$ is the angle between $\bp$ and $\overline{T_xR_x}$, and $\theta^{rx}$ is the angle beеween $\bq$ and $\overline{T_xR_x}$. The electric field generated by the transmitting antenna is propagated in the $S_{\bp}$ plane and attenuated according to $\theta^{tx}$. However, only the portion of the electric field projected to the $S_{\bq}$ plane can contribute to the voltage induction at the receiving antenna, and that portion is further attenuated according to angle $\theta^{rx}$. It can be seen that the mutual orientation of the communicating antennas has a big impact on the voltage induced at the receiving antenna.

The \textbf{key idea} of our solution is to reconstruct the orientation of a transmitting antenna based on the voltage measurements by exploring the above relation. For a BS with a massive MIMO antenna array, the voltages induced from the uplink signals can be measured on the distributed antenna elements. The spatially diversified voltage measurements allow reconstructing the orientation of the transmitting antenna. 

Based on the Hertzian dipole antenna model with the length of $d$ \cite{Orfanidis}, we derive the closed-form equations for the reconstruction of antenna orientation in the single-path \ac{LoS} scenario and then extend the reconstruction methodology to the multipath propagation scenario. 



\subsection{Single-path LoS case}
Let $r$ be the Euclidean distance between the two communicating antennas. As illustrated in Fig.\ref{fig:effective_length}, we use $\be_r^{tx}$ to represent the unit vector for the direction of wave propagation from $T_x$ to $R_x$, and $\be_{\theta}^{tx}$ to represent the unit vector for signal polarization that is always perpendicular to $\be_r^{tx}$ \cite{Orfanidis}. In the same way we have $\be_r^{rx}$ and $\be_{\theta}^{rx}$ for the receiving antenna. Since we consider the dipole antenna model, the observed electric field from angle $\theta^{tx}$ is oscillating within plane $S_{\bp}$. It means that the observed electric field can be considered as if it is transmitted from an antenna $\bp(\theta^{tx})$ with orientation $\be_{\theta}^{tx}$ and length $d\sin{\theta^{tx}}$. In Fig. \ref{fig:effective_length}, $\bp(\theta^{tx})$ is represented by a red line in plane $S_{\bp}$. In the antenna theory, $\bp(\theta^{tx}) = d \sin{\theta^{tx}} \be_{\theta}^{tx}$ is called as the effective length of the transmitting antenna at $T_x$.
Let $\bE(R_x)$ be 
the electric field oscillating in $S_{\bp}$ near the receiving antenna. $\bE(R_x)$ can be defined as
\begin{align}
	\bE(R_x) =  \frac{j \kappa \eta}{4 \pi r} \, \mathrm{I_{in}}e^{-j\kappa r}\bp(\theta^{tx}) 
	=E(r)d \sin{\theta^{tx}} \be_{\theta}^{tx},
	\label{eq:electric_field}  
\end{align}
where $E(r) =  \frac{j \kappa \eta}{4 \pi r} \, \mathrm{I_{in}}e^{-j\kappa r}$ is the scalar part of the electric field measured in $\text{Volts}/\text{meter}^2$, and it is a function of the propagation distance $r$ between the two communicating antennas and the amplitude of the input current to the transmitting antenna $\mathrm{I_{in}}$; $\kappa = \omega/c$ is the wavenumber, $\eta$ is the characteristic impedance of air \cite{Orfanidis}.

Since the receiving antenna can receive the electric field oscillating within plane $S_{\bq}$, only the portion of the electric field projected from $S_{\bp}$ to plane $S_{\bq}$ can be received by the receiving antenna and contribute to voltage induction. The projection of the electric field from $S_{\bp}$  into $S_{\bq}$ is the scalar product of $\bE(R_x)$ with $\be_{\theta}^{rx}$. Due to the reception angle $\theta^{rx}$, the maximum energy reception is further restricted by the effective length of the receiving antenna $\bq(\theta^{rx})=d \sin{\theta^{rx}} \be_{\theta}^{rx}$, which is represented by a green line in Fig.\ref{fig:effective_length}. Hence, the voltage induced at the receiving antenna, denoted by $V$, can be calculated by 
\begin{align}
	V =& \bE(R_x) \cdot \bq(\theta^{rx})\nonumber\\
	=& d^2E(r) \,\sin{\theta^{tx}}\sin{\theta^{rx}} \,(\be_{\theta}^{tx} \cdot \be_{\theta}^{rx}).
	\label{eq:scalar_product}  
\end{align}

\noindent \textbf{Remarks:} It can be seen from Eqn. (\ref{eq:scalar_product}) that no voltage can be induced when $\sin\theta^{tx} = 0$ and (or) $\sin\theta^{rx} = 0$, thereby zeroing out the signal at the receiving antenna. The maximum amplitude for the induced voltage over a given distance $r$ can be obtained if the two antennas are in the same plane and both $\sin{\theta^{tx}}$ and $\sin{\theta^{rx}}$ are equal to $1$ or $-1$. Hence, the mutual orientations and locations have a direct impact on the measured voltage through distance $r$, observation and reception angles $\theta^{tx}$ and $\theta^{rx}$.

The unit vectors $\be_{\theta}^{tx}$ and $\be_{\theta}^{rx}$ can be expressed through $\bp$ and $\bq$, respectively:
\begin{align}
	\be_{\theta}^{tx} = \mPr^{tx} \,\frac{1}{\sin{\theta^{tx}}}\, \bp, \label{eq:e_thetatx}\\ \be_{\theta}^{rx} = \mPr^{rx}\,\frac{1}{\sin{\theta^{rx}}}\,\bq,
	\label{eq:e_thetarx} 
\end{align}
where $\mPr^{tx}\! = \be_r^{tx} (\be_r^{tx})^T\! -\! I$ and $\mPr^{rx}\! = \be_r^{rx} (\be_r^{rx})^T\! -\! I$ are projection matrices that project vectors to $\be_{\theta}^{tx}$ and $\be_{\theta}^{rx}$, respectively. The details on how Eqn. (\ref{eq:e_thetatx}) and Eqn. (\ref{eq:e_thetarx}) are derived are given in the appendix. By substituting $\be_{\theta}^{tx}$ and $\be_{\theta}^{rx}$ in Eqn. (\ref{eq:scalar_product}), we have  
\begin{equation}
	V \! =  d^2 E(r)\! \left(\mPr^{rx} \bq \cdot \mPr^{tx} \bp \right)\! =\bq^T \,\mPr^{rx} d^2 E(r) \mPr^{tx} \, \bp,
	\label{eq:voltage_upd}
\end{equation}
since scalar product $a\cdot b = a^Tb$ and $(\mPr^{rx})^T = \mPr^{rx}$. 

Let $V_n$ be the voltage measured from the $n$-th antenna, and $\mPath_n =  \bq_n^T  \mPr^{rx}_n d^2 E(r_n)  \mPr^{tx}_n$ where $r_n$ is the distance between the transmitting antenna at the UE and the $n$-th antenna at the BS. The notation "Path" is chosen because it represents the transformation that a signal is experiencing during the propagation from the transmitting antenna to the receiving antenna. For a Massive MIMO antenna array with $N$ elements,  Eqn. (\ref{eq:voltage_upd}) can be rewritten as follows:
\begin{equation}
	\spvec{V_1;V_2;\vdots; V_N} = \spvec{\mPath_1 \, ; \mPath_2; \vdots; \mPath_N}\bp.
	\label{eq:voltage_MIMO}
\end{equation}







It can be seen from the system of equations (\ref{eq:voltage_MIMO}) that the left side of the system consists of the real measurements on voltages from  the Massive MIMO antenna array, whereas the right side consists of the reconstructed voltages based on the location $T_x$ and orientation $\bp$ of the transmitting antenna. Hence, the problem to find both the location and orientation of the transmitting antenna can be formulated as the following minimization problem:
\begin{equation}
	\min\limits_{T_x, \bp} \| \bV - \boldsymbol{\mPath} \, \bp \|^2,
	\label{eq:nonlin}
\end{equation}
where $\boldsymbol{\mPath}\! = \!(\mPath_1^T,..,\mPath_N^T)^T$ and $\boldsymbol{V} \!=\! (V_1,..,V_N)^T$. Since both the location and orientation of each receiving antenna at the BS are known, the unknown parameters in (\ref{eq:nonlin}) include: 3 parameters for the location of the transmitting antenna and another 3 parameters for its orientation. Theoretically, both the location and the orientation can be obtained if $N \geq 6$. However, it is worth noting that both $\mPr^{tx}$ and $\mPr^{rx}$ are nonlinear functions of the location of the transmitting antenna. Hence, problem (\ref{eq:nonlin}) becomes a nonlinear optimization problem, which is much harder to solve than linear programs. In practice, the problem (\ref{eq:nonlin}) can be solved in two stages. The first stage is to find the location. Our previous work in \cite{My4} demonstrates the feasibility to achieve decimeter-level accuracy in localizing a UE. Once the location is known, matrix $\boldsymbol{\mPath}$ becomes known and the only unknown in (\ref{eq:voltage_MIMO}) is $\bp$. In this case, the orientation $\bp$ can be found using the standard least squares method as follows:
\begin{equation}
	\Tilde{\bp} = \mathrm{Re}\{(\boldsymbol{\mPath}^T \boldsymbol{\mPath})^{-1}\boldsymbol{\mPath}^T \boldsymbol{V}\}.
	\label{eq:least_squares}
\end{equation}

The solution given in (\ref{eq:least_squares}) is the closed-form solution for the reconstruction of the antenna orientation in the case of LoS propagation. To the best of our knowledge, this is the first time a closed-form solution for antenna orientation reconstruction is derived. 

\subsection{Multipath Case}
In a multipath propagation environment, an antenna can receive a number of copies of the transmitted signal due to signal reflection from reflecting objects. We use the well-known ray tracing approach for modeling multipath propagation, by which the orientation of the transmitting antenna can be explicitly tracked during reflection. In this paper, we do not consider reflections with two or more bounces because, in most practical cases, the energy of a transmitted signal drops sharply after the second reflection according to the Fresnel coefficients of reflection \cite{Rapoport}.

\begin{figure}[!ht]
	\centering
	\includegraphics[width=2in]{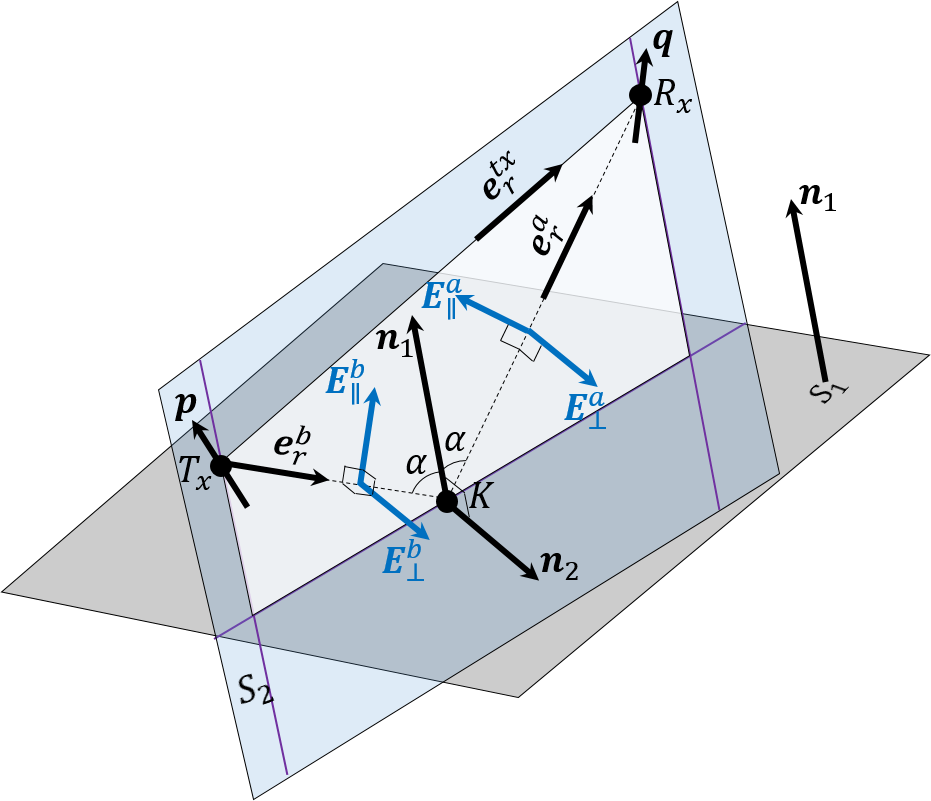}
	\caption{Definition of vectors in the case of reflection.}
	\label{fig:reflection}
\end{figure}
As illustrated in Fig. \ref{fig:reflection}, a transmitted signal is reflected from surface $S_1$ with normal vector $\bn_1$. According to the Law of Reflection, the reflection is proceeding in $S_2$ with normal vector $\bn_2$, and $K$ is the reflection point. 
We use superscript ``$b$" to indicate vectors corresponding to the signal \textit{before} reflection and superscript ``$a$" for the vectors corresponding to the signal \textit{after} reflection. For a LoS path, $\be_r^{tx} = -\be_r^{rx}$. For a NLoS path, $\be_r^{tx} = \be_r^{b}$ and $\be_r^{rx} = -\be_r^{a}$. 

The main challenge for antenna orientation reconstruction in the multipath case is to take into account the transformation of the electric field vector Eqn. (\ref{eq:electric_field}) during reflection. The electric field $\bE^b(K)$ at point $K$ before the reflection can be decomposed into the following two components that transform differently due to the physical properties of the reflecting surface: \textit{(1)} the perpendicular component $\bE_{\bot}^{b}(K)$ that is perpendicular to $S_2$, and \textit{(2)} the parallel component $\bE_{\parallel}^{b}(K)$ that is within $S_2$. We have $\bE_{\bot}^{b}(K)=(\bE^b(K) \cdot \bn_2 ) \bn_2$, and  $\bE_{\parallel}^{b}(K)=\bE^b(K) - \bE_{\bot}^{b}(K)$. Based on the property of the scalar product that $(a\cdot b)c = bc^T a$, the perpendicular component can be transformed as $\bE_{\bot}^{b}(K)=(\bn_2 \bn_2^T) \bE^b(K)$. According to the electric field vector given in Eqn. (\ref{eq:electric_field}) and  Eqn. (\ref{eq:e_thetatx}), the two components at point $K$ before reflection can be computed as follows:
\begin{align}
	\bE_{\bot}^{b}(K) &= d E(r^{K}) (\bn_2 \bn_2^T) \mPr^{b} \bp , \\
	\bE_{\parallel}^{b}(K) &= d E(r^{K}) (I - \bn_2 \bn_2^T) \mPr^{b} \bp, \label{eq:e_parr}
\end{align}
where $r^{K}$ is the distance from $T_x$ to reflection point $K$, and $\mPr^{b} = \be_r^b(\be_r^b)^T - I$ is the projection matrix as in Eqn. (\ref{eq:e_thetatx}).

After the reflection, the perpendicular component attenuates with Fresnel reflection coefficient $\Gamma_{\bot}(\alpha)$ \cite{METIS} where $\alpha$ is the angle of incident. The parallel component rotates in plane $S_2$ clockwise with an angle $\pi - 2\alpha$ to become perpendicular to the propagation direction $\be_r^{a}$ \cite{My2} and attenuates with Fresnel reflection coefficient $\Gamma_{\parallel}(\alpha)$ \cite{METIS}. Hence, the two components after reflection can be represented as follows:
\begin{align}
	\bE_{\bot}^{a}(K) &= \Gamma_{\bot}(\alpha)\, \bE_{\bot}^{b}(K) \\
	& = \Gamma_{\bot}(\alpha) \,d E(r^{K})  (\bn_2 \bn_2^T) \mPr^{b} \bp, \nonumber \\
	\bE_{\parallel}^{a}(K) &= \Gamma_{\parallel}(\alpha) \, W(\bn_2, \pi - 2\alpha) \bE_{\parallel}^{b}(K) \label{eq:e_parr_arr_parallel}\\
	&= \Gamma_{\parallel}(\alpha) \, W(\bn_2, \pi - 2\alpha) d E(r^{K}) (I - \bn_2 \bn_2^T) \mPr^{b} \bp, \nonumber
\end{align}
where $W(\bn_2, \pi - 2\alpha)$ is the rotation matrix that rotates vectors around the normal vector $\bn_2$ with an angle $\pi- 2\alpha$. The main observation is that, in case of NLoS propagation in addition to the propagation attenuation, the electric field experiences additional attenuation caused by the reflection phenomenon. The Fresnel coefficients $\Gamma_{\bot}(\alpha)$ and $\Gamma_{\parallel}(\alpha)$ depend on angle of incident $\alpha$ and the physical properties of the reflecting surface \cite{METIS}.


The transformed electric field defined by $\bE_{\bot}^{a}(K)$ and $\bE_{\parallel}^{a}(K)$ will experience further attenuation when propagating in the direction $\be_r^{a}$ from the reflection point $K$ to the receiver point $R_x$. The portion of the electric field received by the receiving antenna is restricted by the effective length of the receiving antenna, which can be expressed as $\bq(\theta_r^{a}) = d \mPr^{a}\bq$ based on the definition of the effective length and Eqn. (\ref{eq:e_thetarx}).  In the same way as that in the LoS path case, the voltage induced by the electric field propagating along an NLoS can be computed as follows:
\begin{equation}
	V^{NLoS} = \bq^T \mPr^{a}\bE^a(R_x) = \bq^T \mNLoS \,\bp,
	\label{eq:vnlos}
\end{equation}
where $\mPr^{a}\! = \be_r^{a} (\be_r^{a})^T\! -\! I$, $\bE^a(R_x) \,=\, \bE^a_{\bot}(R_x) + \bE^a_{\parallel}(R_x)$, and 
$\mNLoS$ is the electric field transformation matrix defined as follows:
\begin{align}
	\mNLoS &= d^2 \mPr^{a}  E(r^{NLoS})\bigl[ \Gamma_{\bot}(\alpha)(\bn_2 \bn_2^T)\, +\nonumber\\
	&+ \Gamma_{\parallel}(\!\alpha\!) W(\!\bn_2, \pi - 2\alpha\!) (\!I - \bn_2 \bn_2^T\!) \bigr]\mPr^{b}. 
	\label{eq:mNLoS}
\end{align}
Here $r^{NLoS}$ is the total covered distance of the NLoS path.

For the LoS path, its transformation matrix is $\mLoS =  d^2\mPr^{rx} E(r^{LoS}) \,\mPr^{tx}$, and Eqn. (\ref{eq:voltage_upd}) can be written as:
\begin{align}
	V^{LoS} =\bq^T \,\mLoS \,\bp. 
	\label{eq:vlos}
\end{align}

Assume the multipath signal propagation has $L$ NLoS paths. For each NLoS path and the LoS path, the receiving antenna has its vector of effective length. Consequently, the total voltage produced on the receiving antenna can be represented as follows:
\begin{align}
	V & = \bq^T \left[ \mLoS + \sum_{l=1}^{L} \mNLoS_l \right] \bp = \nonumber\\
	& = V^{LoS} + \sum_{l=1}^{L} V^{NLoS}_{l},
	\label{eq:hertzian_transformation_multipath}
\end{align}
where $\mNLoS_l$ is the transformation matrix for the $l$-th NLoS path. Let $\mPath_n =  \bq_n^T \, \bigl[ \mLoS_n + \sum_{l=1}^{L} \mNLoS_{ln} \bigr]$ where $n \in [1, .., N]$. Both the location and orientation of the transmitting antenna can be obtained by solving Problem (\ref{eq:nonlin}) using the same approach as for the LoS case.

\section{Full Downlink Channel Reconstruction}\label{sec:channel_reconstruction}



\subsection{Channel Modeling}



Let us consider the case where the transmitting antennas of the \ac{UE} transmit \ac{UL} signals simultaneously but use different radio resource blocks \cite{LTE-book}. Hence, they don't interfere with each other at the reception side. At the physical level, signals are transmitted via the emission of electromagnetic waves from a transmitting antenna. The control of output electric field defined in Eqn. (\ref{eq:electric_field}) is done by controlling the input current in the time domain $\mI(t) = \mI_{in} \sum_{m = 1}^{N_{\sym}} \sym_m(t) e^{j \omega t}$, where $N_{\sym}$ is the number of transmitting symbols, $\sym_m(t)$ is nonzero in period $[(m-1)\Dt, m \Dt]$ where $\Dt$ is the system's sampling duration, and $\mI_{in}$ is the amplitude of the input current to the transmitting antenna. The \ac{LoS} observation of the electric field at any point $X$, denoted by $\bE(t, X)$, is
\begin{align}
    \bE(t, X) =&  \frac{j \kappa \eta }{4 \pi r} \mI(t - r/c) \bp(\theta^{tx}) = \nonumber \\
     =& \mI_{in} \frac{j \kappa \eta }{4 \pi r} \sum_{m = 1}^{N_{\sym}}\! \sym_m(t\! -\! r/c) e^{j \omega (t-r/c)} \bp(\theta^{tx}), \nonumber
\end{align}
where $r$ is the distance between the antenna and the observation point. At the receiving antenna with orientation $\bq$, this electric field induces voltage $V(t) = \bE(t,X)\cdot \bq(\theta^{rx})$ according to Eqn. (\ref{eq:scalar_product}). By processing the measured voltage, the \ac{BS} can reconstruct the transmitted symbols $\sym_m$ by removing the carrier wave $e^{j\omega t}$ and then estimating channel to equalize the distorted symbols. Through the procedure of \ac{CE} based on the channel estimation reference symbols \cite{LTE-book}, the downlink channel can be represented as follows:
\begin{equation}
    H =  \mI_{in} \frac{j \kappa \eta}{4 \pi r} e^{-j \kappa r} \bp(\theta^{tx})\cdot\bq(\theta^{rx}) = \mI_{in} \mu \kappa  e^{-j \kappa r},
    \label{eq:single_channel}
\end{equation}
where $\mu = \frac{j \eta}{4 \pi r} \bq(\theta^{rx}) \cdot \bp(\theta^{tx})$ can be considered as a complex-valued channel attenuation coefficient. In accordance with Eqn.(\ref{eq:scalar_product}), channel given in Eqn.(\ref{eq:single_channel}) has volt unit. 

In mobility scenarios, the modeling of the downlink channel becomes much more complicated due to the Doppler effect since an arbitrary movement in 3D space causes different velocities on each antenna. Hence, 
in a multipath propagation environment, the signals sent from different antennas can experience different Doppler shifts depending on the propagation path and receiving antenna \cite{Bill_Kihei}:
\begin{equation}
    \nu^{k}_{ln} = (\bv^k \cdot \be^k_{ln}) \frac{\omega}{c} = D^{k}_{ln} \kappa,
    \label{eq:doppler_effect}
\end{equation}
where $\bv^k$ is the velocity vector of the $k$-th antenna, $\kappa = \omega/c$ is the wavenumber, and $\be^k_{ln}$ is the unit length radius vector that indicates the direction from the $k$-th transmitting antenna to the $n$-th receiving antenna through the $l$-th path. Hence, in a dynamic multipath propagation environment, the channel for the signal that travels from the $k$-th transmitting antenna to the $n$-th receiving antenna can be modeled based on Eqn. (\ref{eq:single_channel}) and Eqn. (\ref{eq:doppler_effect}) as:
\begin{equation}
    H_n^k = \mI_{in}\sum_{l=0}^L \mu_{ln}^k \kappa \, e^{-j\kappa r_{ln}^k} \, e^{-j \nu^{k}_{ln}(r_{ln}^k/c)},
    \label{eq:multi_channel}
\end{equation}
where $L$ is the number of propagation paths. 
To enable channel reconstruction, the multipath propagation parameters $\{\mu_{ln}^k, r_{ln}^k,\nu^{k}_{ln}\}$ for each propagation path $l$ from the $k$-th transmitting antenna to the $n$-th receiving antenna have to be identified. The total number of unknown multipath propagation parameters becomes larger than the number of measured channels since there is only $m$ ($m<M$) transmitting antennas.  To find all the parameters, we leverage the \ac{OFDM} nature to increase the number of measurements \cite{RIMAX_Transaction}. Consequently, the channel at subcarrier $f_i = \omega + i \Delta f$ can be represented as follows:
\begin{equation}
    H_n^k(f_i)\! = \!\mI_{in}\sum_{l=0}^{L} \mu^k_{ln}\kappa_i \,e^{ -j \kappa_i \, r_{ln}^k} \,e^{-j D_{ln}^k \kappa_i (r_{ln}^k/c)},
    \label{eq:channel_estimation}
\end{equation}
where $r_{ln}^k$ and $D_{ln}^k$ are the frequency independent parameters of distance and Doppler shift, respectively. Channel attenuation coefficients $\mu_{ln}^k$ are frequency dependent; $\kappa_i = f_i/c$, $f_i \in \mathcal{F}_k$ and $\mathcal{F}_k$ is the subset of subcarriers that is allocated for the $k$-th transmitting antenna, and $\cup_{k=1}^m \mathcal{F}_k$ is the total given radio resource.

\noindent\textbf{Observation:} 
The BS can reconstruct the downlink channels once the multipath propagation parameters $\{\mu_{ln}^k, r_{ln}^k, D^{k}_{ln}\}$ are obtained from Eqn. (\ref{eq:channel_estimation}) \cite{MIT-Vasisht1}. While the parameters for the transmitting antennas can be easily inferred based on the measurements of the UL signals, it is challenging to infer them for the non-transmitting antennas. The following explains how to infer these parameters. 

\subsection{Parameter Estimation for Transmitting Antennas}
The propagation parameters $\{\mu_{ln}^k, r_{ln}^k, D^{k}_{ln}\}$ can be estimated by solving the following optimization problem:
\begin{equation}
    \min\limits_{\{\mu_{ln}^k, r_{ln}^k, D^{k}_{ln}\}} \sum \limits_{i=1}^{N_s^k}\sum \limits_{n=1}^{N} \sum \limits_{k=1}^{m} \Big\|{H_n^k}'(f_i)\! - \!H_n^k(f_i)\Big\|^2,
    \label{eq:optimization}
\end{equation}
where $N_s^k$ is the total number of subcarriers in $\mathcal{F}_k$. Such kind of optimization problem can be solved using one of the standard optimization methods such as Levenberg-Marquardt, SAGE \cite{SAGE1}, or RIMAX in a more complex propagation model \cite{RIMAX_Transaction} with dense multipath components. 

The main obstacle in estimating the propagation parameters is that $r_{ln}^k$ and $D_{ln}^k$ have to be estimated as one parameter $r_{ln}^k +D_{ln}^k(r_{ln}^k/c)$ since they cannot be separated from the exponential function $e^{-j \kappa_i (r_{ln}^k +D_{ln}^k(r_{ln}^k/c))}$. Hence, by solving Problem (\ref{eq:optimization}), we can get $\mu_{ln}^k$ and $r_{ln}^k +D_{ln}^k(r_{ln}^k/c)$. To further separate $r_{ln}^k$ and $D_{ln}^k$, we perform parameters estimation twice with a time gap $\tau$. In fact, \ac{UL} channel estimation is performed twice every millisecond in LTE \cite{LTE-book}, which gives the required time diversity in estimated parameters. We assume that the change of the \ac{UE}'s position between two \ac{CE}s is negligible. For $100$ m/s, the position change is $5$ cm. Suppose  the estimation for the first \ac{CE} is $est_1 = r_{ln}^k +D_{ln}^k(r_{ln}^k/c)$, and at the second is $est_2 = r_{ln}^k +D_{ln}^k(r_{ln}^k/c - \tau)$. The difference between the two estimations is $est_1 - est_2 = D_{ln}^k \tau$. Since $\tau$ is known, we can extract $D_{ln}^k$ from the difference of two consecutive \ac{CE}s. Once $D_{ln}^k$ is extracted, $r_{ln}^k$ can also be obtained.

Based on the \ac{SWP} model, the locations of the transmitting antennas and their images can be found from the extracted $r_{ln}^k$ \cite{My4}. As illustrated in Fig. \ref{fig:UE} (a), both the location of the transmitting antenna $T_{x}^k$ and its image $\mathrm{Im}_l(T_{x}^k)$ can be estimated using the solution presented in \cite{My4}, based on which the reflecting plane $S_l$ can be determined because it has to go through the middle of the segment $[T_{x}^k, \mathrm{Im}_l(T_{x}^k)]$ and be perpendicular to this segment. Now, to calculate the NLoS transformation matrices in Eqn. (\ref{eq:mNLoS}), ARDI needs to calculate the angle of incidence $\alpha_{ln}^k$, which can be obtained based on the location of the receiving antenna and the reflecting plane $S_l$. Hence, by doing the same operations for each pair of antennas $(T_{x}^k, R^n_x)$ and each $l$-th path, ARDI can calculate the incidence angles $\alpha_{ln}^k$ for each NLoS path.

The \ac{CE} given in Eqn. (\ref{eq:multi_channel}) can be represented in a convenient form for orientation reconstruction as follows:
\begin{equation}
    H_n^k = \bq_n^T\biggl[ \mLoS \mathrm{D}_n^k + \sum_{l=1}^L \mNLoS\mathrm{D}_{ln}^k \biggr] \bp_k,
\end{equation}
where the multipath transformation matrices incorporate the estimated Doppler shifts $\mLoS \mathrm{D}_n^k = \mLoS_n^k \,e^{-j D_{0n}^k \kappa (r_{0n}^{cg}/c)}$ and $\mNLoS\mathrm{D}_{ln}^k =\mNLoS_{ln}^k\, e^{-j D_{ln}^k \kappa (r_{ln}^{cg}/c)}$.

\subsection{Parameter Estimation for Non-transmitting Antennas}

\begin{figure}
	\centering
	\includegraphics[width=3.3in]{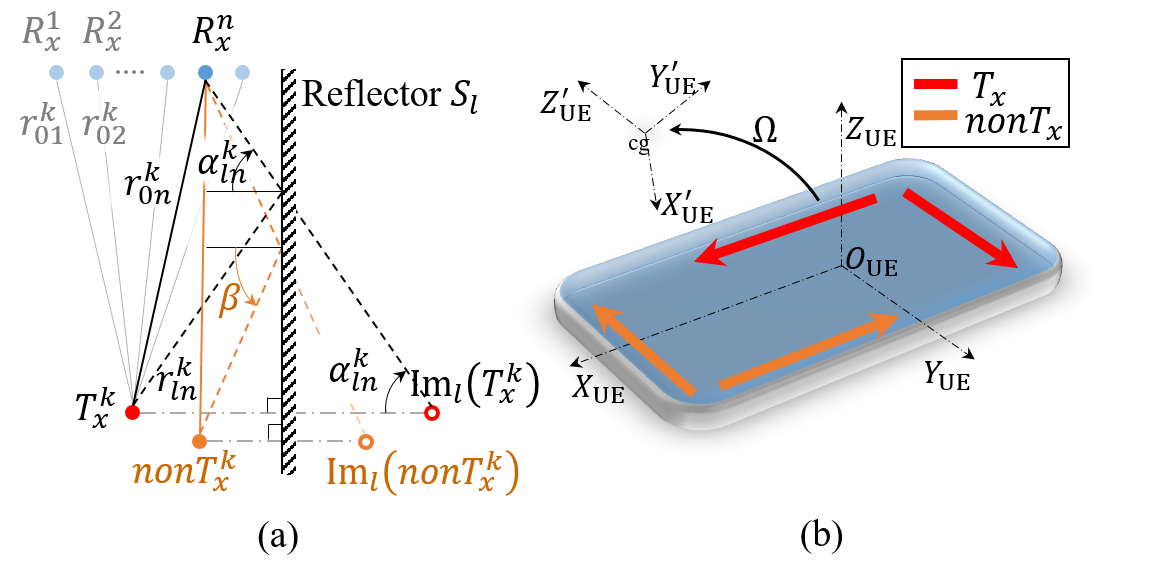}
	\caption{(a) Derivation of the reflection plane from a transmitting antenna location and its image location. (b) Mobility of UE, its antennas' orientation and location.}
	\label{fig:UE}
\end{figure}

As illustrated in Fig. \ref{fig:UE} (b), we assume at least two of the $m$ transmitting antennas are not parallel. During the \ac{CE} procedure, ARDI can obtain the coordinates and orientations of the $m$ transmitting antennas using the method introduced in Section \ref{sec:antenna_orientation}, from which the orientation of the \ac{UE}, denoted by $\Omega$, can be estimated. Suppose the layout the antennas in UE is a priori knowledge based on the UE design, both the positions and orientations of the non-transmitting antennas can then be calculated based on their relative positions/orientations to the transmitting antennas. 

\textbf{Inference of $r_{ln}^k$:} For the LoS path, $r_{ln}^k$ can be easily calculated based on the locations of the transmitting and receiving antennas. For the NLoS path, we use the example given in Fig. \ref{fig:UE} (a) to explain the inference of $r_{ln}^k$. Suppose $nonT_{x}^k$ represents the location for a  non-transmitting antenna. Based on the reflecting plane $S_l$, the image of $nonT_{x}^k$, denoted by $\mathrm{Im}_l(nonT_{x}^k)$, can be calculated since the location of $nonT_{x}^k$ is known. Then $r_{ln}^k$ can be obtained by calculating the distance from the image to the receiving antenna, and the incident angle $\alpha_{ln}^k$ can also be obtained. 

\textbf{Inference of $\mu_{ln}^k$:} 
Once $r_{ln}^k$ and the orientations of the communicating antennas have been found, $\mu_{ln}^k$ can be inferred since $\mu_{ln}^k = \frac{j \eta}{4 \pi r_{ln}^k} \bq(\theta^{rx}) \cdot \bp(\theta^{tx})$. 


\textbf{Inference of $D^{k}_{ln}$:}
To infer the Doppler effects, ARDI needs to obtain at least two measurements of the location and orientation of the \ac{UE}. Using these measurements, it calculates the speed and the angular velocity as follows: 
\begin{equation*}
\bv_{cg} = \frac{\mathrm{UE}_{cg}(t_2)-\mathrm{UE}_{cg}(t_1)}{t_2 - t_1},\,\Dot{\Omega} = \frac{\Omega(t_2)-\Omega(t_1)}{t_2 - t_1},
\end{equation*}
where $\mathrm{UE}_{cg}$ is the averaged position of the transmitting antennas, which roughly coincides with the UE's center of gravity, $t_1$ and $t_2$ are moments when the location and orientation measurements have been obtained. Consequently, speeds of antennas are inferred as $\bv^k = \bv_{cg} + \Dot{\Omega} T_x^k$, and Doppler shifts are calculated according to Eqn. (\ref{eq:doppler_effect}). 

\begin{figure}[!ht]
	\centering
	\includegraphics[width=3in, trim=3.5cm 1cm 3.5cm 1cm, clip]{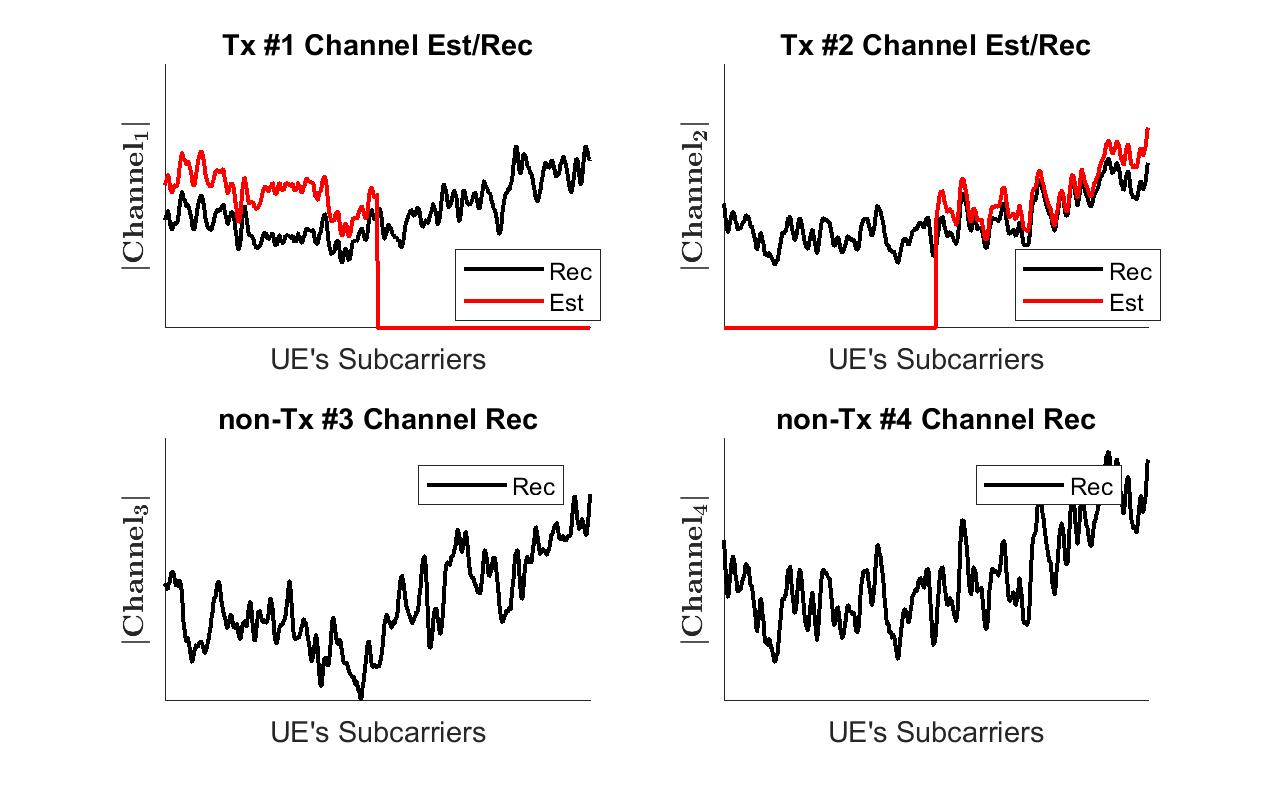}
	\caption{Example with estimated and inferred channels for a \ac{UE} with four antennas two of them are transmitting in LTE TDD transmission mode.}
	\label{fig:ChEstRec}
\end{figure}

After inference the multipath propagation parameters $\{\mu_{ln}^k, r_{ln}^k, D^{k}_{ln}\}$, ARDI reconstructs the \ac{DL} channels using Eqn. (\ref{eq:channel_estimation}) for non-transmitting antennas. Fig. \ref{fig:ChEstRec} gives an example with 2 transmitting antennas and 2 non-transmitting antennas, where the red lines represent the measured channels and the black lines represent the reconstructed channels. For uplink transmission, the full radio resource is equally allocated to the two transmitting antennas. For DL channel inference, each antenna occupies the full radio resource. 

\subsection{Feasibility for FDD and TDD modes}
It can be seen that our scheme reconstructs the full DL channel based on the environmental parameters: propagation paths, mobility, location, and orientation of the UE's antennas. Based on the incomplete uplink channel measurements, ARDI infers these parameters and reconstructs the channels separately for all UE antennas. This means that ARDI creates a separate model of the environment for each antenna where the antenna is the only transmitting antenna. In this way, without interference from other antennas, each antenna can occupy the full radio resource and transmit signals from its estimated position with estimated orientation.
This feature makes ARDI capable of inferring \ac{DL} channel in both  \ac{TDD} and \ac{FDD} transmission modes since there is no difference in frequency choice from the inference perspective.



\section{Performance Evaluation}\label{sec:simulation}
In this section, we present the results for simulation-based evaluation of ARDI based on realistic UE movement.


\subsection{Simulation Setup}
\begin{figure}[!ht]
	\centering
	\includegraphics[width=2in, trim=75cm 10cm 60cm 30cm, clip]{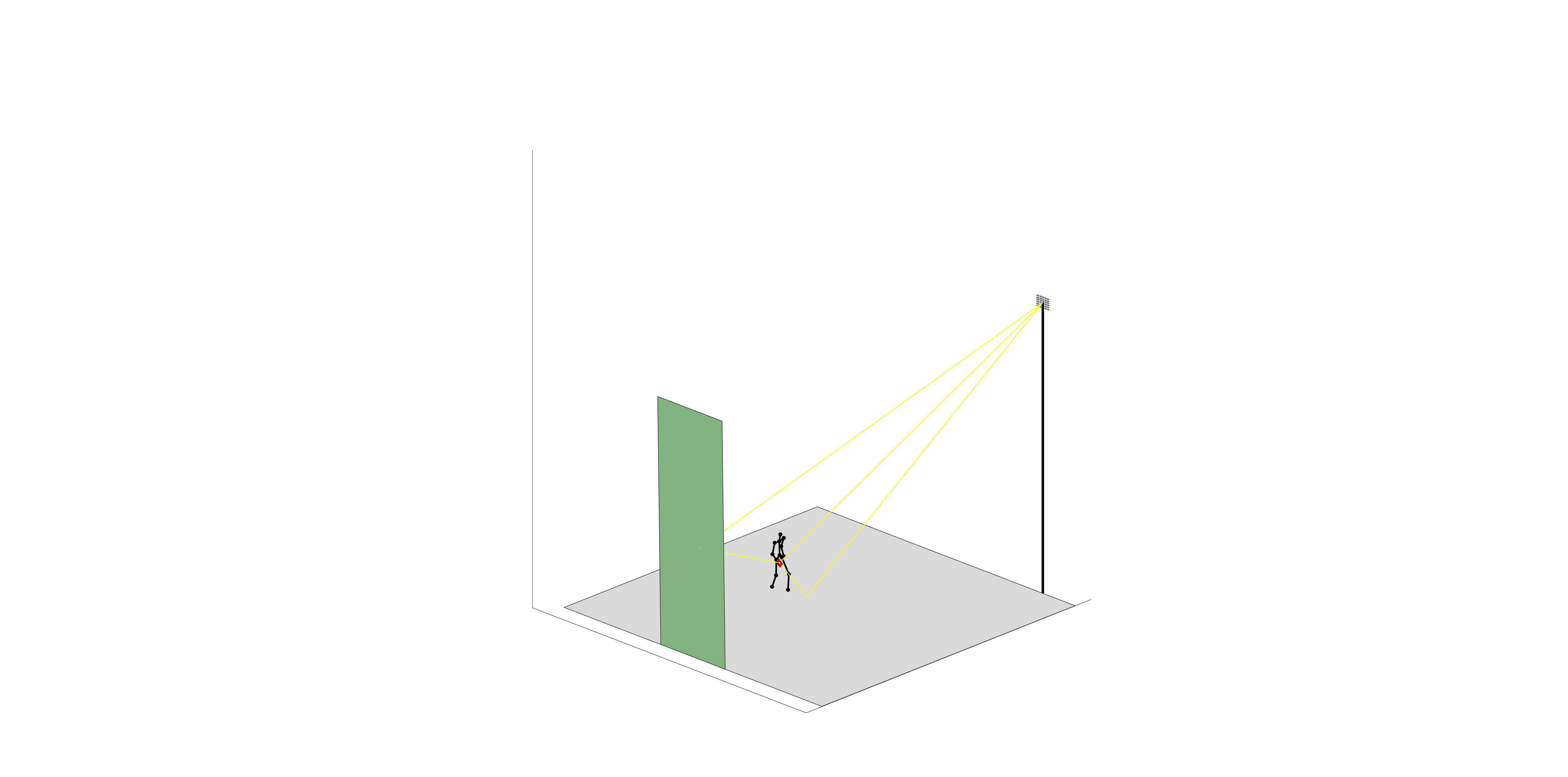}
	\caption{Simulation environment for signal propagation. }
	\label{fig:human}
\end{figure}

As shown in Fig. \ref{fig:human}, the UE is kept in the human's right hand, and we use the well-known eigenwalker model \cite{human} to model the movement of a human body. 
The yellow lines represent the propagation paths. During a random walk, in addition to the ground reflection, six reflecting planes are simulated by randomly positioning and arbitrarily orienting them in the 3D space. However, we depict only one reflector in the example since adding the other planes, and propagation paths can make the plot messy. The signals are propagated according to the \ac{SWP} model and reflected based on the law of reflection.

\subsubsection{Signal specification}
we simulate the typical LTE communication with carrier frequency $\omega = 2.4$ GHz. The \ac{UE} is a cellphone with four antennas, and only two of them can transmit. The LTE signals from two transmitting antennas occupy $100$ resource blocks, with $12$ subcarriers in each and $15$ kHz separation between subcarriers \cite{LTE-book}. In total, $1200$ subcarriers are equally shared by the two transmitting antennas.  Additive Gaussian noise with zero mean value is applied at the \ac{BS} side. The intensity of the noise is defined by the \ac{SNR} relative to the strength of the \ac{LoS} signal. Even in the case where the \ac{LoS} path is blocked, the noise intensity is calculated relative to the \ac{LoS} signal as if it has been delivered to the \ac{BS}. 

\subsubsection{Geometry specification}
as illustrated in Fig. \ref{fig:human}, the \ac{BS} has a planar antenna array that consists of $256$ antennas, $16$ rows in horizontal and $16$ columns in vertical directions. Antennas are half wavelength separated in both directions. Orientations of antennas are set in the way that each next antenna has alternated orientation \{East, North, Up\}. The location of the \ac{BS} is fixed, and the height is $20$ meters above the ground. The \ac{UE} is modeled as a red rectangular polygon with $120\times70$ millimeters in length and width. Four antennas are located on the edges of the polygon in two parallel pairs, as shown in Fig. \ref{fig:UE} (b).  The height of the human is $1.65$ meters. The distance between the \ac{BS} and \ac{UE} varies from $50$ to $100$ meters.
The average moving speed is set to $5$ kilometers per hour. Due to a realistic motion of the human's model and 3D motion of the UE, the speeds of antennas on the \ac{UE} differ from each other. This creates different Doppler effects for different antennas. 

\subsubsection{Physical parameters of the environment} The ground is assumed to be bricked. The reflectors are made from concrete. Relative permittivity and conductivity parameters of these materials are taken from the Material properties Table in \cite{METIS}. The air attenuation is considered as free space attenuation. 

We use the Levenberg-Marquardt algorithm based on the standard Matlab function $\mathrm{lsqnonlin}$ to extract the parameters from Eqn. (\ref{eq:optimization}) \cite{My4}. Once the algorithm converges, we feed the extracted results to ARDI to analyze the accuracy of antenna orientation reconstruction and channel reconstruction. We consider two propagation scenarios: \textit{scenario 1} - multipath propagation with \ac{LoS} when the \ac{LoS} path is observable; \textit{scenario 2} - multipath propagation without \ac{LoS} when the \ac{LoS} path and the path reflected from the ground are blocked. 

\begin{figure*}[!ht]
    \centering
    \subfloat[]{{\includegraphics[width=2.1in]{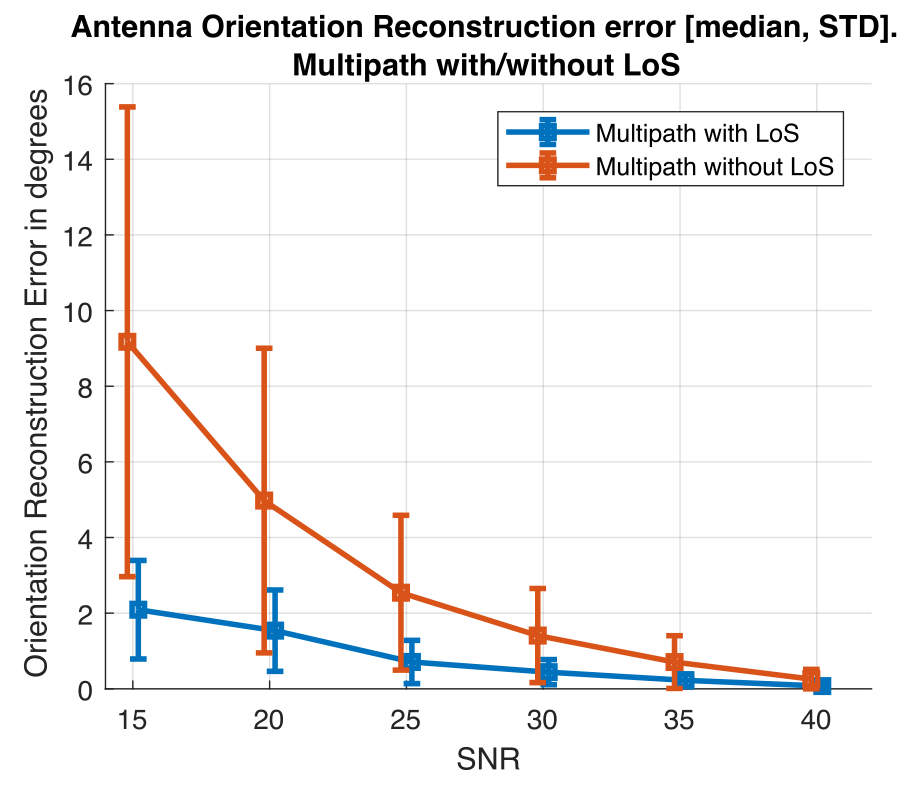} }}%
    \quad
    \subfloat[]{{\includegraphics[width=1.9in]{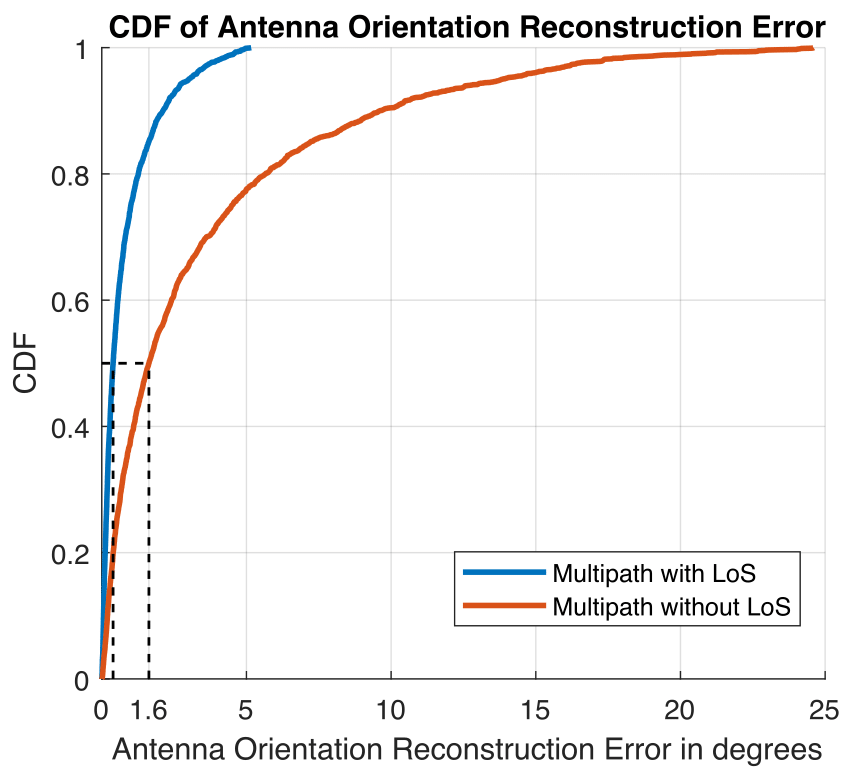} }}%
    \quad
    \subfloat[]{{\includegraphics[width=2.5in]{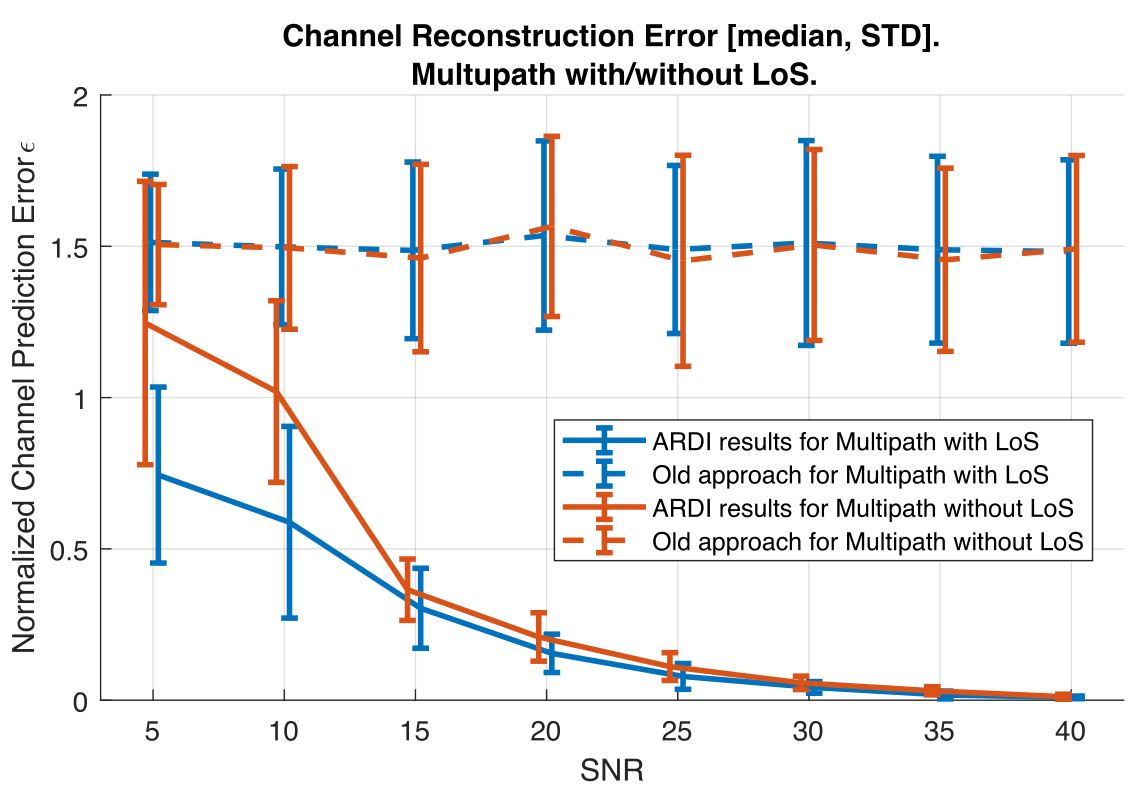} }}%
    \caption{(a) Accuracy of antenna orientation estimation; (b) CDF of antenna orientation estimation; (c)  accuracy of DL channel reconstruction.}%
    \label{fig:simulation_results}%
\end{figure*}



\subsection{Results on antenna orientation reconstruction} 
Fig. \ref{fig:simulation_results} (a) plots the median and standard deviation of the antenna orientation reconstruction error measured in degrees under different setting of \ac{SNR}. Each point is calculated based on the results of $500$ iterations. It is well seen that the antenna orientation reconstruction error tends to converge to zero with the increase of \ac{SNR} in both scenarios. 
The accuracy of antenna orientation reconstruction is worse in \textit{scenario 2}. 
This is an expected result caused by the blockage of the strong \ac{LoS} path, which leads to a less accurate estimation of propagation parameters. 
In addition, transformation matrix $\bmPath$ in Eqn. (\ref{eq:hertzian_transformation_multipath}) accumulates less observation, which additionally degrades the overall reconstruction performance. However, such kind of accuracy is enough to achieve good accuracy on \ac{DL} channel reconstruction, which will be demonstrated in the following subsection. Fig. \ref{fig:simulation_results} (b) shows the \ac{CDF} of the orientation reconstruction errors for the considered \ac{SNR}s. The median estimation error is less than $2^{\circ}$ even for \textit{scenario 2}. This capability can be used in different types of application such as the elimination of \ac{DL} channel feedback in beamforming, human motion tracking, localization refining, etc.



\subsection{Results on full downlink channel reconstruction}

In this section, we evaluate the performance of ARDI for \ac{DL} channel reconstruction in \ac{TDD}  mode. In the simulation, we first measure the channels separately for each pair of communicating antennas $\bf{\overline{H_n^k}}$ = $({H_n^k}(f_1),.., {H_n^k}(f_{N_s}))^T$ where $n \in [1,...,256]$, $k\in[1,..., 4]$,  $N_s = 1200$, and ${H_n^k(f_i)}$ is defined in Eqn. (\ref{eq:multi_channel}). Then we crop the measured channels for the two transmitting antennas $({\bf{\overline{H_{n}^{1}}}},{\bf{\overline{H_{n}^{2}}}})$ by providing each antenna with $600$ subcarriers. As illustrated in Fig. \ref{fig:ChEstRec}, the first antenna occupies the first $600$ subcarriers, and the second occupies the rest part of subcarriers. After the crop, ARDI extracts propagation parameters $\{\mu_{ln}^k, r_{ln}^k, D_{ln}^k\}$ from the cropped channels. 
Then, ARDI performs antenna orientation reconstruction and DL channel inference for the non-transmitting antennas $(\bf{{H_n^3}},\bf{{H_n^4}})$, which are further compared with the measured channels $({\bf{\overline{H_{n}^{3}}}},{\bf{\overline{H_{n}^{4}}}})$. We compare ARDI with an ``old" approach in which the \ac{BS} just uses the measured channels for the transmitting antennas as predictions for the corresponding parallel non-transmitting ones. In other words, the measured channels $\bf{\overline{H_{n}^{1}}}$ are used to predict channels for non-transmitting antenna \#$3$, and $\bf{\overline{H_{n}^{2}}}$ is used to predict channels for non-transmitting antenna \#$4$.

We aim to analyze the difference between the reconstructed channel and the measured channel in terms of both amplitude and phase. However, the absolute difference between the measured and reconstructed channels is not representative since the absolute values of the channels for different distances may differ by the orders of magnitudes. The channel differences for locations that are far from the BS can be much smaller than the differences for locations that are close to the BS. Due to this, we propose to use the following metric, which can be considered as a normalized difference between reconstructed and measured channels: 
\begin{equation}
    \epsilon_n^k = \frac{\|{\bf{\overline{H_{n}^{k}}}} - \bf{H_n^k}\|}{\| {\bf{\overline{H_{n}^{k}}}} \|}, n \in [1,..,256], k \in [3,4],
    \label{eq:metric}
\end{equation}
where ${\bf{\overline{H_{n}^{k}}}}$ is the measured channel and $\bf{H_n^k}$ is the reconstructed channel. It can be seen that the closer the reconstructed channel to the measured channel, the smaller the value of $\epsilon$. This metric takes into account not only the correlation of the channels but also the similarity of amplitudes and complex phases of the channels. For example, the channels may be well correlated with coefficient of correlation $\rho({\bf{\overline{H_{n}^{k}}}},\bf{H_n^k}) \approx 1$ while ${\bf{H_n^k}} = \beta {\bf{\overline{H_{n}^{k}}}}$ where $\beta$ can be any complex number. It is also critical to make $\beta$ close to one especially in a precoding procedure \cite{LTE-book}, which means the difference $\|{\bf{\overline{H_{n}^{k}}}} - \bf{H_n^k}\|$ has to be close to zero. Let us consider the case where the amplitudes of
${\bf{\overline{H_{n}^{k}}}}$ and $\bf{H_n^k}$ are quite similar, which is possible when localization is performed accurately. We can define $\bf{H_n^k}$ = $\Phi {\bf{\overline{H_{n}^{k}}}}$ where $\Phi$ is a unitary matrix in the vector space $\mathbb{C}^{N_s}$. The metric can be re-written as  $\epsilon_n^k = \| (I_{N_s} - \Phi){\bf{\overline{H_{n}^{k}}}} \|/\| {\bf{\overline{H_{n}^{k}}}} \|$. Since the amplitudes are similar, the difference on phase will dominate the reconstruction error. Hence, if $\Phi = I_{N_s}$, $\epsilon_n^k = 0$, which indicates that the channel is accurately reconstructed. If $\Phi = -I_{N_s}$, $\epsilon_n^k = 2$, which is the worst case.  Consequently, if the amplitudes of the measured and the reconstructed channels are similar,  $\epsilon_n^k \in [0, 2]$. For the case where the amplitudes are quite different, it is possible when there is a big error on localization. For such a case, the difference on amplitude can dominate the reconstruction error, and $\epsilon_n^k$ can be any positive value fenced from zero. 


To examine the overall channel reconstruction performance, we run $100$ iteration. For each iteration we calculate median and standard deviation of $\{\epsilon_n^k\}$ $n\in [1,..256]$, $k\in[3,4]$, and then average all the obtained medians and standard deviations. Fig. \ref{fig:simulation_results} (c) shows the average medians and standard deviations under different settings on SNR. 
It can be seen that the reconstructed channels converge to the measured channels in both scenarios with the increase of SNR. One observation from this figure is that the ``old" approach does not give good channel prediction for any \ac{SNR}. Note, if the metric is far above zero, the reconstructed channel is less likely related to the real channel. 
Another interesting observation can be seen from the standard deviation. For lower \ac{SNR}s, ARDI has a larger standard deviation than the ``old" approach. This is because, for lower \ac{SNR}s, ARDI reconstructs DL channel inaccurately because of the large errors in estimating the propagation parameters and antenna orientation and locations. Instead of reconstructing the channel near the \ac{UE}, ARDI reconstructs the channel for a distant place from the \ac{UE}. In such kind of situations, the value of $\epsilon$ can significantly increase and can be higher than $2$. However, in most cases, ARDI still can correctly reconstruct DL channel, and this is why the median is lower than that in the ``old" approach. As expected, the performance of \ac{DL} channel reconstruction is slightly weaker in \textit{Scenario 2}. For higher \ac{SNR}s starting from $15$ dB, ARDI performs similarly in both cases. Based on the obtained results we can conclude that ARDI is capable of reconstructing the full \ac{DL} channel with reasonable accuracy for \ac{SNR}s higher than $15$ dB. This revolutionary ability of ARDI can become very helpful in the reconstruction of the full Massive \ac{MIMO} channel from the incomplete channel measurements. 



\section{Conclusion}\label{sec:conclusion}
In this paper, we introduce ARDI, a scheme that is capable of reconstructing the full downlink channel in Massive MIMO systems from incomplete UL channel measurements in both FDD and TDD communication modes. ARDI enables the increase of Massive MIMO channel capacity without further growth of the number of transmitting antennas. Our work can have implication for other types of wireless communication systems such as WiFi and mmWave networks since the same physical principles are used in all of them. Further development of this research lays in the extension of the antenna orientation reconstruction method towards realistic antenna models and experimental validation.

\section{Appendix}
Following the rules of vector product, it is well seen that the unit vector $\be_{\theta}^{tx}$ is defined by the antenna orientation $\bp$ and the radius vector $\be_{r}^{tx}$ as follows:
\begin{equation}
	\be_{\theta}^{tx} = \frac{ (\bp \times \be_{r}^{tx})\times \be_{r}^{tx}}{\|(\bp \times \be_{r}^{tx})\times \be_{r}^{tx}\|}.
	\label{eq:vect_product}
\end{equation}
The numerator can be rewritten as $\left((\be_r^{tx} (\be_r^{tx})^T) - I_3\right) \bp$ where $I_3$ is the identity matrix in 3D space, and the denominator as $\sin{\theta^{tx}}$ since $(\bp \times \be_{r}^{tx}) = \sin{\theta^{tx}} \be_{\vphi}^{tx}$, where $\be_{\vphi}^{tx}$ compliments $\be_{r}^{tx}, \be_{\theta}^{tx}$ to a right-handed triple $(\be_{r}^{tx},\be_{\theta}^{tx}, \be_{\vphi}^{tx})$, and $\be_{\vphi}^{tx}\times \be_{r}^{tx} = \be_{\theta}^{tx}$. We define a matrix $\mPr^{tx} = \left((\be_r^{tx} (\be_r^{tx})^T) - I\right)$. The meaning of this matrix is that it projects orientation vector $\bp$ to $\be_{\theta}^{tx}$. Hence, we can get Eqn. (\ref{eq:e_thetatx}) and Eqn. (\ref{eq:e_thetarx}).


\bibliographystyle{IEEEtran}
\bibliography{AF_bibliography}

\begin{thebibliography}{10}
\providecommand{\url}[1]{#1}
\csname url@samestyle\endcsname
\providecommand{\newblock}{\relax}
\providecommand{\bibinfo}[2]{#2}
\providecommand{\BIBentrySTDinterwordspacing}{\spaceskip=0pt\relax}
\providecommand{\BIBentryALTinterwordstretchfactor}{4}
\providecommand{\BIBentryALTinterwordspacing}{\spaceskip=\fontdimen2\font plus
\BIBentryALTinterwordstretchfactor\fontdimen3\font minus
  \fontdimen4\font\relax}
\providecommand{\BIBforeignlanguage}[2]{{%
\expandafter\ifx\csname l@#1\endcsname\relax
\typeout{** WARNING: IEEEtran.bst: No hyphenation pattern has been}%
\typeout{** loaded for the language `#1'. Using the pattern for}%
\typeout{** the default language instead.}%
\else
\language=\csname l@#1\endcsname
\fi
#2}}
\providecommand{\BIBdecl}{\relax}
\BIBdecl

\bibitem{Ericsson}
Ericsson, ``{Massive MIMO increasing capacity and spectral efficiency},''
  \url{https://www.ericsson.com/en/news/2018/1/massive-mimo-highlight},
  [Online; accessed 26-January-2018].

\bibitem{Huawei}
Huawei, ``{Massive MIMO is the future of wireless networks},''
  \url{https://www.huawei.com/en/about-huawei/publications/winwin-magazine/28/massive-mimo-2016},
  [Online; accessed 21-June-2017].

\bibitem{Facebook}
Facebook, ``{Introducing Facebook’s new terrestrial connectivity systems —
  Terragraph and Project ARIES},''
  \url{https://code.fb.com/connectivity/introducing-facebook-s-new-terrestrial-connectivity-systems-terragraph-and-project-aries/},
  [Online; accessed 13-April-2016].

\bibitem{Samsung}
Samsung, ``{Specifications | Samsung Galaxy Note9 – The Official Samsung
  Galaxy Site},''
  \url{https://www.samsung.com/global/galaxy/galaxy-note9/specs/}.

\bibitem{Qualcomm}
``{Qualcomm unveils first mmWave 5G antennas for smartphones},''
  \url{https://www.theverge.com/2018/7/23/17596746/qualcomm-mmwave-5g-antenna-smartphones-qtm052-networking-speeds-size}.

\bibitem{Marzetta}
J.~Jose, A.~Ashikhmin, T.~L. Marzetta, and S.~Vishwanath, ``Pilot contamination
  and precoding in multi-cell tdd systems,'' \emph{IEEE Transactions on
  Wireless Communications}, vol.~10, no.~8, pp. 2640--2651, August 2011.

\bibitem{MIT-Vasisht1}
D.~Vasisht, S.~Kumar, H.~Rahul, and D.~Katabi, ``Eliminating channel feedback
  in next-generation cellular networks,'' in \emph{Proceedings of the 2016
  Conference on ACM SIGCOMM 2016 Conference}, ser. SIGCOMM '16.\hskip 1em plus
  0.5em minus 0.4em\relax New York, NY, USA: ACM, 2016, pp. 398--411.

\bibitem{arXiv_DLCH_reconstruction}
Y.~{Han}, T.-H. {Hsu}, C.-K. {Wen}, K.-K. {Wong}, and S.~{Jin}, ``{Efficient
  Downlink Channel Reconstruction for FDD Multi-Antenna Systems},'' \emph{ArXiv
  e-prints}, May 2018.

\bibitem{Lund_RIMAX}
X.~Li, K.~Batstone, K.~Åstrom, M.~Oskarsson, C.~Gustafson, and F.~Tufvesson,
  ``Robust phase-based positioning using massive mimo with limited bandwidth,''
  in \emph{2017 IEEE 28th Annual International Symposium on Personal, Indoor,
  and Mobile Radio Communications (PIMRC)}, Oct 2017, pp. 1--7.

\bibitem{RIMAX_Transaction}
J.~Salmi, A.~Richter, and V.~Koivunen, ``Detection and tracking of mimo
  propagation path parameters using state-space approach,'' \emph{IEEE
  Transactions on Signal Processing}, vol.~57, no.~4, pp. 1538--1550, April
  2009.

\bibitem{My4}
A.~Fedorov, H.~Zhang, and Y.~Chen, ``User localization using random access
  channel signals in lte networks with massive mimo,'' in \emph{2018 27th
  International Conference on Computer Communication and Networks (ICCCN)},
  July 2018, pp. 1--9.

\bibitem{SAGE1}
B.~H. Fleury, M.~Tschudin, R.~Heddergott, D.~Dahlhaus, and K.~I. Pedersen,
  ``Channel parameter estimation in mobile radio environments using the sage
  algorithm,'' \emph{IEEE Journal on Selected Areas in Communications},
  vol.~17, no.~3, pp. 434--450, March 1999.

\bibitem{orient1}
A.~Shahmansoori, G.~E. Garcia, G.~Destino, G.~Seco-Granados, and H.~Wymeersch,
  ``Position and orientation estimation through millimeter-wave mimo in 5g
  systems,'' \emph{IEEE Transactions on Wireless Communications}, vol.~17,
  no.~3, pp. 1822--1835, March 2018.

\bibitem{Orfanidis}
S.~J. Orfanidis, \emph{{Electromagnetic Waves and Antennas}}.\hskip 1em plus
  0.5em minus 0.4em\relax Rutgers University New Brunswick, NJ, 2014.

\bibitem{Rapoport}
O.~Landron, M.~Feuerstein, and T.~Rappaport, ``A comparison of theoretical and
  empirical reflection coefficients for typical exterior wall surfaces in a
  mobile radio environment,'' \emph{IEEE Transactions on Antennas and
  Propagation}, vol.~44, no.~3, pp. 341--351, 1996.

\bibitem{METIS}
L.~Raschkowski, P.~Ky\"{o}sti, K.~Kusume, and T.~J\"{a}ms\"{a}, ``{METIS
  Channel Models},'' METIS, Tech. Rep., 2015.

\bibitem{My2}
A.~Fedorov, H.~Zhang, and Y.~Chen, ``Geometry-based modeling and simulation of
  3d multipath propagation channel with realistic spatial characteristics,'' in
  \emph{2017 IEEE International Conference on Communications (ICC)}, May 2017,
  pp. 1--6.

\bibitem{LTE-book}
S.~Sesia, I.~Toufik, and M.~Baker, \emph{{LTE - The UMTS Long Term Evolution:
  From Theory to Practice}}.\hskip 1em plus 0.5em minus 0.4em\relax Chichester:
  John Wiley \& Sons Ltd, 2011.

\bibitem{Bill_Kihei}
B.~Kihei, J.~A. Copeland, and Y.~Chang, ``Cepstral analysis for classifying car
  collisions in los/nlos vehicle-to-vehicle networks,'' in \emph{GLOBECOM 2017
  - 2017 IEEE Global Communications Conference}, Dec 2017, pp. 1--6.

\bibitem{human}
N.~Troje, ``The little difference: Fourier based gender classification from
  biological motion,'' \emph{Dynamic Perception}, 01 2002.

\end{thebibliography}

\end{document}